\documentclass[authoryear,preprint,12pt,3p,times,onecolumn]{elsarticle}
\usepackage{amsmath,amssymb,graphicx,pifont,float,afterpage}

\usepackage{lineno}

\renewcommand{\d}{\mathrm{d}}

\journal{Process Safety and Environmental Protection }
\begin{document}

\begin{frontmatter}

\title{Thermal instability and runaway criteria: \\the dangers of disregarding dynamics} 
\author{ R. Ball\corref{cor1}}
\address{Mathematical Sciences Institute, The Australian National University\\ Canberrra ACT 0200 Australia }
\ead{Rowena.Ball@anu.edu.au}
\cortext[cor1]{Corresponding author}

\author{ B.F. Gray\corref{cor1}}
\address{School of Mathematics and Statistics F07,
University of Sydney NSW 2006
Australia  }
\ead{bfgrayaus@hotmail.com}

\begin{abstract}
Two exemplary exothermic processes,  synthesis of nitroglycerine in a continuous stirred tank reactor (CSTR) and synthesis of the explosive RDX in a CSTR, are used to demonstrate the dangers of ignoring the system dynamics when defining criteria for thermal criticality or runaway.  Stability analyses are necessary to prescribe such criteria, and for these systems prove the presence of dangerous oscillatory thermal instability  which cannot be detected using the steady state thermal balances.  
\end{abstract}

\end{frontmatter}
%\linenumbers
 \section{Introduction}
 This short communication is an adjuration for the correct use of mathematical stability analysis in specifying critical runaway conditions for open thermoreactive systems, and is presented in the interests of process  safety. In considering the thermal stability of such a system, operating at  a given steady state, the first and most important question we should ask is  
 \begin{description}
 \item[Question 1.]  Will a small perturbation to the temperature grow uncontrollably, or decay harmlessly back to the steady state? 
 
 \end{description}
 
 Open dissipative dynamical systems in general may become unstable at a turning point (also called a  limit point or a saddle-node bifurcation) or at a Hopf bifurcation to an oscillatory state, and this is the case for open thermoreactive systems too. 
 Some open reacting systems in which thermal oscillations have been observed experimentally since 1969 are listed in Table \ref{table0}. Typically the setup is a continuous stirred tank reactor (CSTR), which is an exemplary open system for observing bistable and oscillatory phenomena as well as being  widely used in industry. 
\begin{table}\caption{\label{table0} Experimental CSTR systems in which thermal oscillations have been observed.} 
 \centerline{\footnotesize
 \begin{tabular}{p{0.6\textwidth}p{0.3\textwidth}}
 \hline\hline
 Reaction & Reference\\
 \hline
 Vapour-phase chlorination of methyl chloride & \cite{Bush:1969}\\
  Hydrolysis of acetyl chloride in acetone-water solvent & \cite{Baccaro:1970}\\
Sodium thiosulfate and hydrogen peroxide reaction in aqueous solution  &  \cite{Chang:1975}\\
Decomposition of hydrogen peroxide by Fe$^{3+}$ in HNO$_3$& \cite{Wirges:1980}\\
 Acid catalysed hydration of 2,3-epoxypropanol-1  to glycerol & \cite{Heemskerk:1980,Vermeulen:1986a,Vermeulen:1986b}\\
 H$_2$SO$_4$ catalysed hydrolysis of acetic anhydride &\cite{Haldar:1991,Jayakumar:2011} \\
 Hydrolysis of methyl isocyanate (modelled, using experimental data) & \cite{Ball:2011}\\
 \hline\hline
 \end{tabular}
 }
 \end{table} 
 
Oscillatory instabilities are also well-understood mathematically.   The mathematical framework for identifying and characterising such instabilities in dynamical systems  was set out by  \cite{Maxwell:1868} and is presented  in many excellent modern texts; rather than cite them  here or reproduce the mathematical details the reader is referred to the expository and resources chapter by \cite{Ball:2007}. This mathematics of dynamical systems has long been accessible to chemical engineers. It  has  been applied to 
  reactive thermal systems by many authors and their results  published widely in the chemical engineering literature since the late 1950s.  Some notable examples are as follows: \cite{Aris:1958}; \cite{Bush:1969};  \cite{Gray:1969}, where stability conditions are worked through in detail, and  the abstract   states explicitly ``The critical condition for an oscillatory stable steady state is of an entirely new type (in explosion theory) not reducible to a tangency condition'';  and     \cite{Gray:1975}. 

The alarm bells ring, therefore, when we see recent works specifying thermal runaway criteria where the the process dynamics and  mathematical stability  have been completely disregarded. Given the potential  of oscillatory instability to cause serious thermal hazard,  initiating thermal runaway, unintended thermal explosions, or worse,  we feel    that it is a matter of urgency to bring the importance of stability analysis to the attention of those sectors of the chemical engineering community which deal with reactive thermal hazards and runaway criteria, and correct certain misconceptions prevalent in the current literature regarding thermal stability criteria.  This is best achieved by use of concrete examples of real systems. The following two published examples  \citep{Lu:2008,Lu:2005} were chosen as cases that graphically illustrate the pitfalls of ignoring CSTR \textit{dynamics},  and where  stability analysis reveals oscillatory thermal instability of dangerous amplitude that cannot be detected using classical (Semenov) ignition theory.  

 \section{Example I: Synthesis of nitroglycerine in a CSTR} \label{s2}

\cite{Lu:2008} developed  CSTR operating criteria for  the esterification of glycerol with nitric acid to produce nitroglycerine,
\begin{equation}\label{r1}
\text{C}_3\text{H}_5(\text{OH})_3 + 3\text{HNO}_3 \overset{\text{H}_2\text{SO}_4}{\longrightarrow} \text{C}_3\text{H}_5(\text{ONO}_2)_3 +3\text{H}_2\text{O}, 
\end{equation} 
 that were claimed to be safe, based on applying classical ignition criteria to a \textit{steady state} thermal balance.   (But see the derivation of CSTR equations given in \ref{B}.) Their mass and enthalpy balances for this system, expressed in \textit{dynamical} form, are 
\begin{align}
V\frac{\text{d}x}{\text{d}t }&= VA\text{e}^{E/RT} (c_\text{G,f}-x)^n(c_\text{N,f}-3x)^m  + F(c_\text{G,0}-c_\text{G,f} -x) \label{e0}\\
V\overline{C}_\text{vol}\frac{\text{d}T}{\text{d}t} &= 
V\left(\left(-\Delta H^\circ\right) + \Delta C_p \left(T-T^\circ\right)\right) A\text{e}^{E/RT} (c_\text{G,f}-x)^n(c_\text{N,f}-3x)^m - \overline{C}_\text{vol}F(T-T_\text{f}) - L(T-T_\text{a}).\label{e1}
\end{align}   
Notation and quantities are defined in  \ref{A}.  (\cite{Lu:2008} took $c_\text{G,0}\equiv c_\text{G,f}$ and $T_\text{f}\equiv T^\circ$.) 

\begin{figure}[ht]
  \centerline{
  \includegraphics[scale=.8]{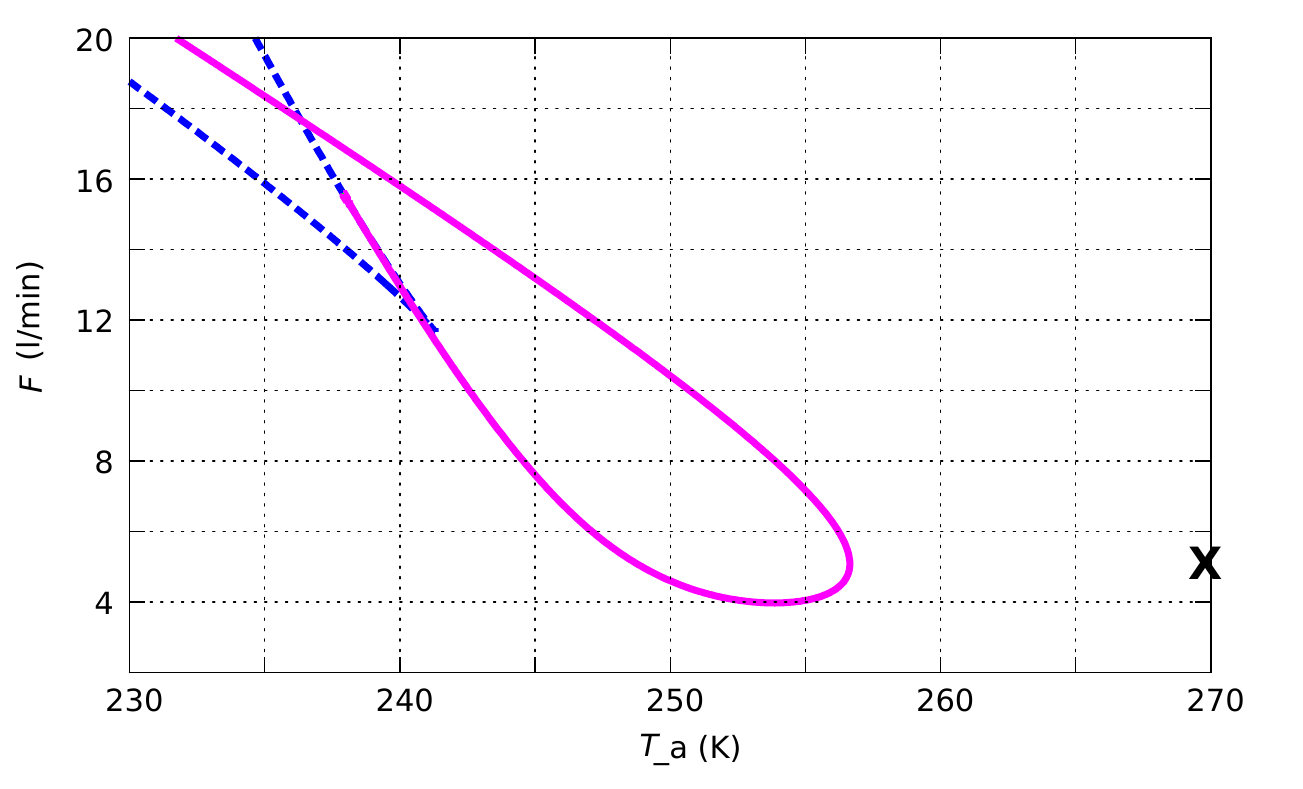}
  }
  \caption{\label{figure1} Loci of Hopf  (solid line, magenta) and saddle-node (dashed line, blue) bifurcations in the system  (\ref{e0}) and (\ref{e1}). The \textsf{\textbf{X}} marks a steady state considered by  \cite{Lu:2008}. $L=80079$\,W/K, $T_\text{f}=273$\,K, $C_\text{vol}=2703$\,J\,K$^{-1}$l$^{-1}$, $\Delta H^\circ=100$\,kJ/mol, $\Delta C_p=-25$\,J\,K$^{-1}$mol$^{-1}$, $n=0.935$, $m=1.117$, $A=9.78\times 10^{22}$(l/mol)$^{1-n-m}$min$^{-1}$, $E=122$\,kJ/mol, $c_\text{G,0}=c_\text{G,f}=2.4904$\,mol/l, $c_\text{N,0}=8.5535$\,mol/l.}
  \end{figure}

Equations (\ref{e0}) and (\ref{e1}) are not amenable to exact analysis, but the stability can be mapped by numerical computation. One such map is rendered in Fig. \ref{figure1}, where the loci of critical points are projected  on the $T_\text{a}$--$F$ parameter plane, and the values of the other system parameters are those given in \cite{Lu:2008}. To obtain the data for the curves in the figure we computed the steady states of Eqs  (\ref{e0}) and (\ref{e1})  for increments in $T_\text{a}$, beginning at $T_\text{a}=230$\,K and with $\Delta T_\text{a}=0.04$\,K, and for $F=14$\,l/min. At each steady state the corresponding linear eigenvalue problem was solved and the points where an eigenvalue changed sign were flagged. The system becomes unstable if such a sign change is positive. These flagged points are the bifurcation points:  if a real eigenvalue becomes positive there is a saddle-node bifurcation, or classical ignition point,  and if the real parts of a complex conjugate pair become positive the system exhibits a Hopf bifurcation and the onset of limit cycle oscillations.  Then, $F$ was allowed to vary as well as $T_\text{a}$ and the loci of the Hopf and saddle-node bifurcations was computed over $T_\text{a}$--$F$. 

  In Fig. \ref{figure1} the solid (magenta) line is the locus of the Hopf bifurcations. Within the region that is almost enclosed by this line, and also near to the upper section of the line, the system will develop self-sustained thermal oscillations, regardless of the initial conditions. In the region enclosed by the dashed (blue) line the system has two stable  steady states, between which is an unstable steady state. 
  
\begin{figure}
  \centerline{
  \includegraphics[scale=1]{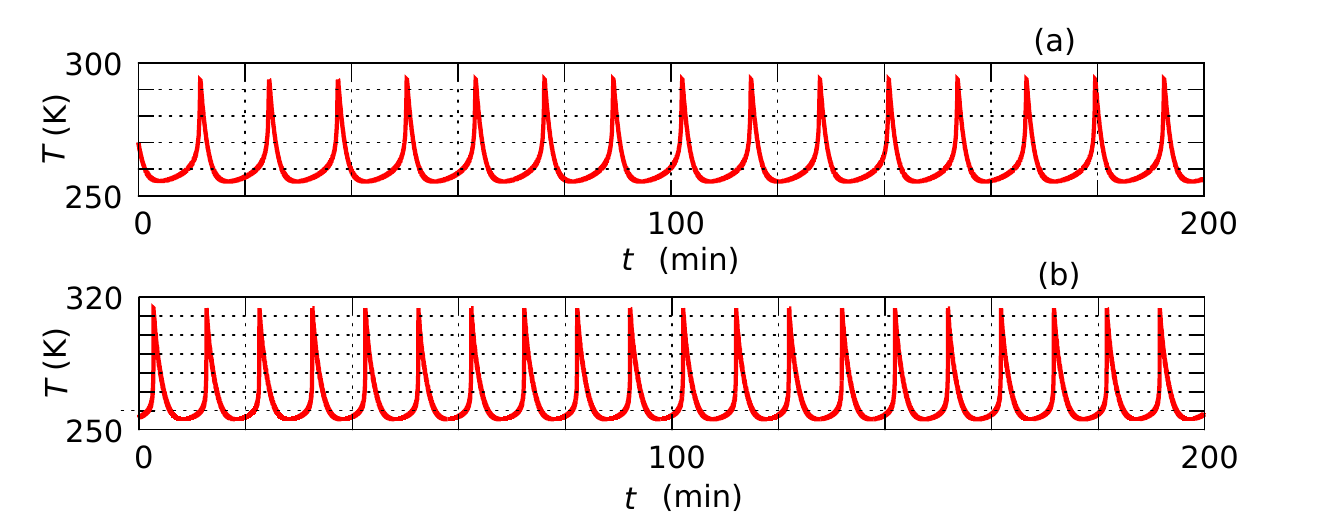}
  }
  \caption{\label{figure2} Computed time series for selected values of $T_\text{a}$ and $F$ within the unstable regime in Fig. \ref{figure1}. (a) $T_\text{a}=250$\,K, $F=5$\,l/min. (b) $T_\text{a}=246$\,K, $F=9$\,l/min. Other parameters as given for Fig. \ref{figure1}. }
  \end{figure}

  The \textsf{\textbf{X}} in Fig. \ref{figure1} marks one of the steady states that was considered by  \cite{Lu:2008}. As can be seen, it is outside the region of oscillatory instability and the region of multiplicity  and a check of the computed data finds the steady state system temperature $T=281$\,K.  But if $T_\text{a}$, the coolant temperature, is reduced below 257\,K (with $F$ fixed) the Hopf bifurcation locus is crossed and thermal oscillations set in. A time series computed at $T_\text{a}=250$\,K is shown in Fig. \ref{figure2}(a); the temperature amplitude maximum of $\sim$295\,K may cause some concern. This amplitude actually increases with increasing $F$ and decreasing $T_\text{a}$: the time series shown in   Fig. \ref{figure2}(b) has a seriously dangerous amplitude maximum. 
  
  We would like to emphasize that the presence of these oscillatory states can only be detected by applying stability analysis correctly to the full dynamical system, Eqs (\ref{e0}) and (\ref{e1}). The steady state thermal balance as used by  \cite{Lu:2008} is blind to oscillatory solutions because it cannot answer, or even ask, Question 1.  At best a steady state thermal balance can give the curve of saddle-node bifurcations in Fig. \ref{figure1}, but since it cannot make any statement about the system stability at such a point it is, strictly speaking,  inadequate even for specifying classical ignition/extinction criteria. Furthermore, a classical ignition point does not automatically infer thermal runaway or criticality. A check of the computed data for $F=14$\,l/min finds the steady state temperature $T=267$\,K at the putative `ignition' point and $T=257$\,K at the `extinction' point. These are very low temperatures for this system, and in this case  the saddle-node locus is not  useful for defining  thermal criticality. 
  
\begin{figure}
  \centerline{
  \includegraphics[scale=1]{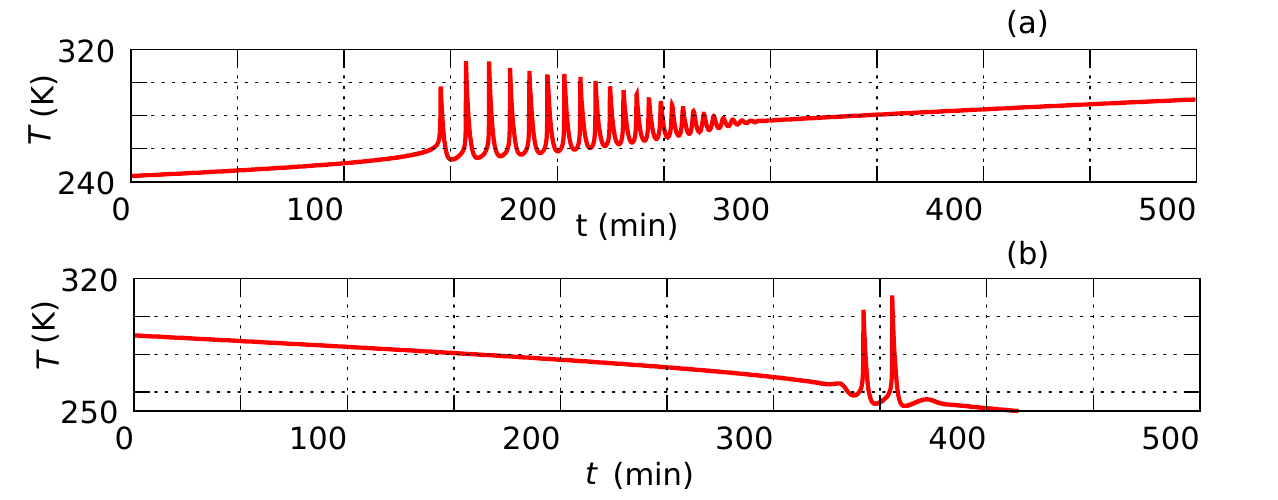}
  }
  \caption{\label{figure3} Computed time series for  $F=8$\,l/min with simulated slow drift or tuning of ambient temperature, (a) $\dot{T}_\text{a}=0.07$\,K/min, (b) $\dot{T}_\text{a} =-0.07$\,K/min.  Other parameters as given for Fig. \ref{figure1}. }
  \end{figure}

  We also point out that the presence of oscillatory instability may have serious implications for start-up and shut-down procedures. Figure \ref{figure3} shows the time series computed for a slow drift or tuning of $T_\text{a}$,  $\dot{T}_\text{a}=\pm 0.07$\,K/min. In both cases the onset of thermal oscillations is violent and the amplitude is dangerously high. 

   \section{Example II: Synthesis of RDX in a CSTR}
  
  Synthesis of  the explosive cyclo-1,3,5-trimethylene-2,4,6-trinitramine, commonly known as RDX or hexogen or cyclonite, by reaction of hexamine with excess nitric acid is exothermic and known to exhibit thermal instability \citep{Luo:2002}. The synthesis is often carried out in 100\% HNO$_3$ at $-20^\circ$C \citep{Singh:1999}. The following overall reaction includes a secondary reaction:
 \begin{equation}\label{reaction1}
2\text{C}_6\text{H}_{12}\text{N}_4 + 26\text{HNO}_3 \rightarrow \underset{\text{RDX}}{\text{C}_3\text{H}_6\text{N}_6\text{O}_6} +9\text{CH}_2{(\text{ONO}_2)}_2
+ 5\text{NH}_4\text{NO}_3 + 3\text{H}_2\text{O}. 
\end{equation}
The primary reaction is  
\begin{equation}\label{reaction2}
\text{C}_6\text{H}_{12}\text{N}_4 + 10\text{HNO}_3 \rightarrow \underset{\text{RDX}}{\text{C}_3\text{H}_6\text{N}_6\text{O}_6} +3\text{CH}_2{(\text{ONO}_2)}_2
+ \text{NH}_4\text{NO}_3 + 3\text{H}_2\text{O}, 
\end{equation}
 This nitration was modelled by  \cite{Lu:2005} as a CSTR process using reaction (\ref{reaction1}). Taking the steady state enthalpy balance they applied Semenov theory  to map the classical ignition and extinction points over the parameter space. 
However,  consideration of the \textit{stability} of the steady states yields a different picture. The CSTR mass and enthalpy balances used by \cite{Lu:2005} (but see \ref{B}), expressed in dynamical form, are 
\begin{align}
V\frac{\text{d}c}{\text{d}t }&= -VA\text{e}^{E/RT} c_\text{N}c^n + F(c_\text{f}-c) \label{e2}\\
V\overline{C}_\text{vol}\frac{\text{d}T}{\text{d}t} &= 
V\left(\left(-\Delta H^\circ\right) + \Delta C_p \left(T-T^\circ\right)\right) A\text{e}^{E/RT}c_\text{N} c^n - \overline{C}_\text{vol}F(T-T_\text{f}) - L(T-T_\text{a}).\label{e3}
\end{align} 
Notation and quantities are defined in  \ref{A}.  

\begin{figure}[ht]
  \centerline{
  \includegraphics[scale=.8]{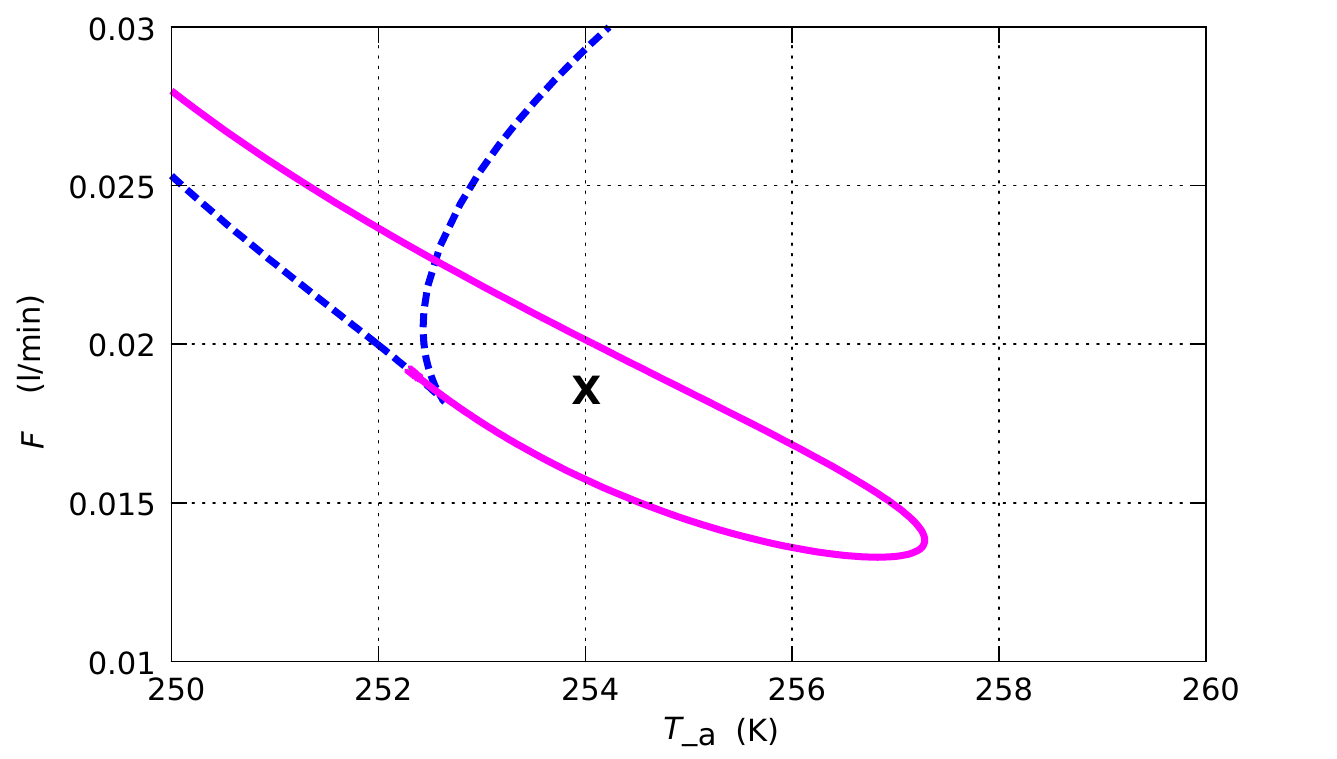}
  }
  \caption{\label{figure4} Loci of Hopf bifurcations (solid line) and saddle-node bifurcations in the system  (\ref{e2}) and (\ref{e3}).  \textsf{\textbf{X}} marks the values of $T_\text{a}$ and $F$ for which the time series in Fig. \ref{figure5} was computed. $L=50$\,W/K, $T_\text{f}=273$\,K, $C_\text{vol}=2940$\,J\,K$^{-1}$l$^{-1}$, $\Delta H=90.938$\,kJ/mol, $\Delta C_p=-322$\,J\,K$^{-1}$mol$^{-1}$, $n=1.28$, $A=1.234\times 10^{6}$(l/mol)$^{n}$min$^{-1}$, $E=47.152$\,kJ/mol, $c_\text{f}=4.805$\,mol/l, $c_\text{N}=21.35$\,mol/l, $T^\circ=298$\,K.}
  \end{figure}

We carried out a numerical stability analysis of this system using a  procedure exactly analogous to that described in section \ref{s2}, and found that it, too,  exhibits oscillatory instability.  In Fig. \ref{figure4} the solid (magenta) line marks the locus of the Hopf bifurcations  dashed (blue) line marks the locus of saddle-node bifurcations, or classical ignition points $F$--$T_\text{a}$, for the given values of the other parameters. The qualitative similarity to Fig. \ref{figure1} may be noted, and the interpretation given in section \ref{s2} may be applied here too. 

\begin{figure}[H]
  \centerline{
  \includegraphics[scale=.8]{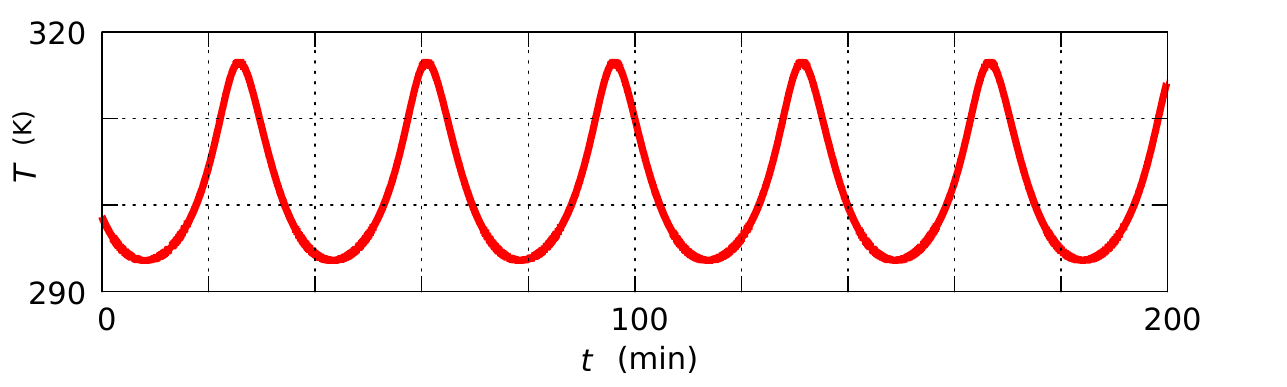}
  }
  \caption{\label{figure5}A time series integrated from Eqs (\ref{e2}) and (\ref{e3}), with  $T_\text{a}=254$\,K, $F=0.0185$\,l/min and other parameters as given for Fig. \ref{figure4}.}
  \end{figure}

The value of $T_\text{a}$ and of $F$  marked by the \textbf{\textsf{X}} in Fig. \ref{figure4} were chosen to compute a time series; this is shown in Fig. \ref{figure5}. The high amplitude of the thermal oscillation, at 317\,K, is of concern: above $\sim$300\,K  exothermic side-reactions are reported to  take over  \citep{Hale:1925}, rendering the system extremely unsafe. 

The standard reaction enthalpy  includes the enthalpy of mixing nitric acid and hexamine. From standard enthalpies of formation  \cite{Luo:2002} found $\Delta H^\circ $(298\,K)$=-90.938$\,kJ/mol hexamine for the overall reaction (\ref{reaction1}), and this is the value we used in the computations. However,   
for the main reaction (\ref{reaction2}) $\Delta H^\circ $(298\,K) $=-153.318$\,kJ/mol hexamine for the reaction only, from standard enthalpies of formation. 
Both of these values are at odds with that measured by \cite{Dunning:1952} using straightforward calorimetry, 
at least 64\,kcal/mol$= 270$\,kJ/mol hexamine at $-35.5^\circ$C in 99\% nitric acid, which gives  $\Delta H^\circ $(298\,K)$=290$\,kJ/mol hexamine; presumably exothermic secondary reactions were involved. In any case, we can expect the oscillatory instabilities to be more violent and to occur over more of the parameter space for a  more exothermic process. 
  
\begin{figure}
  \centerline{
  \includegraphics[scale=.6]{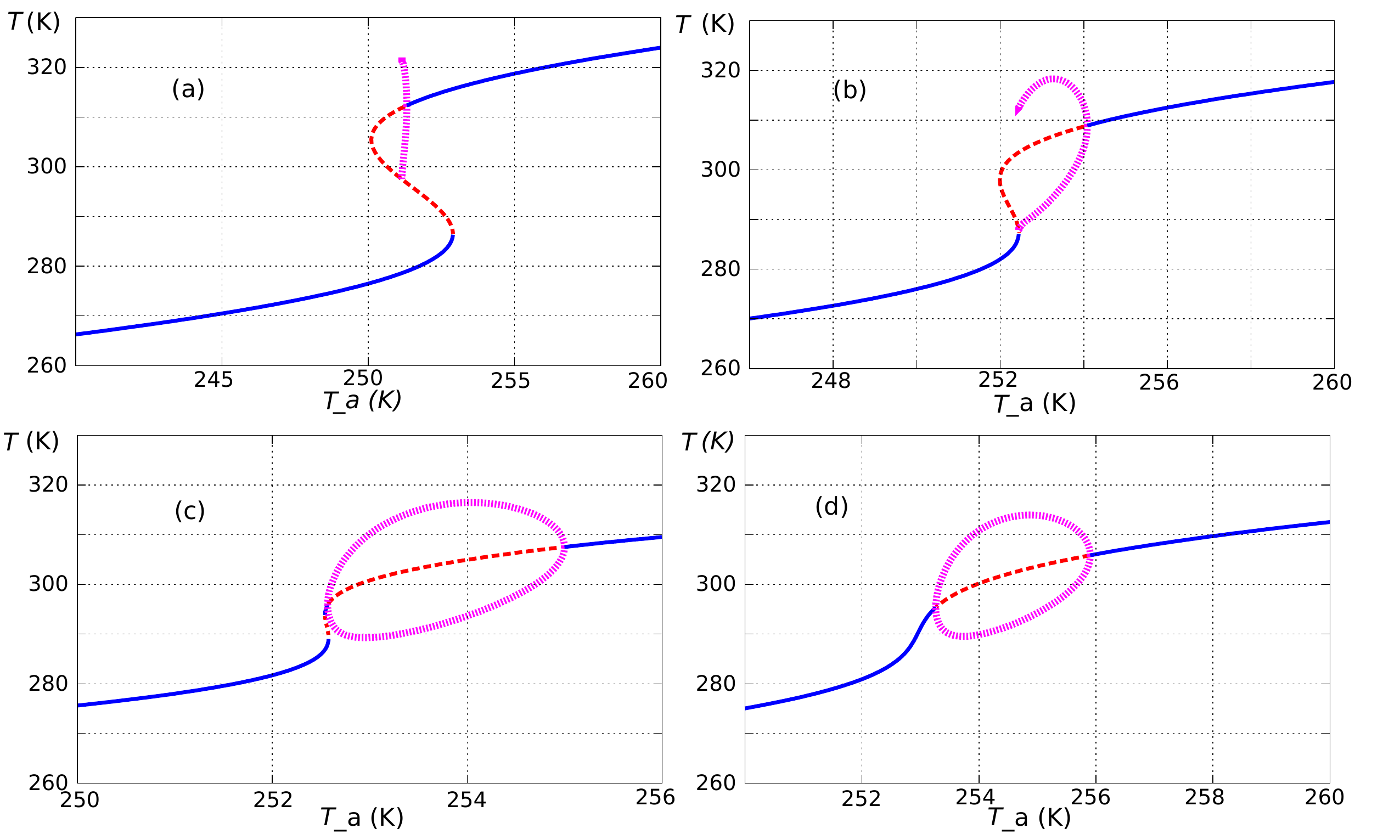}
  }
  \caption{\label{figure6} Stable (solid lines, blue) and unstable (dashed lines, red) steady states and amplitude envelopes of limit cycles (thick dotted lines, magenta)  in the system  (\ref{e2}) and (\ref{e3}).  (a)  $F=0.025$\,l/min, (b) $F=0.020$\,l/min, (c) $F=0.0185$\,l/min, (d) $F=0.017$\,l/min. Other parameters as given for Fig. \ref{figure4}.}
  \end{figure}

It is helpful to study the computed solutions rendered in terms of the temperature $T$ over $T_\text{a}$  in Fig. \ref{figure6}, for selected values of $F$. Stable steady states are marked with solid lines (blue) and unstable steady states are marked with dashed lines (red). Also marked are the maximum and minimum   amplitude envelopes of the limit cycles (which were also computed and analysed for stability), with thick dotted lines (magenta), from their origin at a Hopf bifurcation to their termination at a second Hopf bifurcation (in (c) and (d))  or at a homoclinic bifurcation (in (a) and (b)). These bifurcation diagrams are a rich source of information about the thermal stability of the system. 

In Fig. \ref{figure6}(a) we see that classical ignition occurs at the lower saddle-node bifurcation to a stable steady state on the upper stable solution branch. This does not necessarily represent thermal runaway, although the temperature of $\sim$313\,K may be too high for comfort. The temperature is reduced by quasistatic reduction of $T_\text{a}$ --- but as the Hopf bifurcation is crossed rapidly growing thermal oscillations set in, and at the terminus $T\sim 322$\,K. We also see that extinction does \textit{not} occur at the second saddle-node bifurcation, as assumed by \cite{Lu:2005},  but at this oscillatory terminus (provided thermal runaway has not already set in). 

In Fig. \ref{figure6}(b) the ignition is non-classsical; it occurs from the lower saddle-node bifurcation to a  limit cycle that grows in amplitude with $T_\text{a}$, reaches a maximum, then declines. 

Steady state multiplicity is  barely present at the value of $F$ for which Fig. \ref{figure6}(c) was computed. Non-classical ignition from the lower saddle-node bifurcation occurs to a limit cycle, but extinction in this case is classical, occuring at the upper saddle-node bifurcation. 

The system globally has no multiplicity in Fig. \ref{figure6}(d) and thermal runaway, if it occurs, must be the result of oscillatory behaviour.

  \section{Discussion and summary} 
 
 The two examples treated here provide ample illustration of the dangers inherent in using a steady state thermal balance and ignoring the dynamics when determining thermally safe operating criteria for open thermoreactive processes.  Classical (Semenov) ignition theory was originally applied to explosive solid materials then extended to reacting gases at low pressure, for which it works well.  No reference is made in Semenov theory to oscillatory solutions, because reactant depletion is ignored (the infinite pool approximation) and periodic solutions are automatically forbidden in a system evolving in just one dynamical variable---the temperature.  Semenov theory is insufficient for analysing  open flow systems with reactant depletion,  for which the steady states \textit{and} their stability must be fully characterised. For the simple case of a single first-order (or pseudo first-order) exothermic conversion in a CSTR the global point of onset of multiplicity has been calculated analytically  \citep{Ball:1999} and analysis combined with numerics has characterised the oscillatory states and identified thermal runaway due to the hard onset of oscillatory instability at a subcritical Hopf bifurcation \citep{Ball:1995a,Ball:2011}. For more complicated or more accurately modelled systems numerical eigenvalue analyses can be carried out, as we have done here.  

More broadly, we are concerned that crucial knowledge that was available (and, of course, is still available) to and made good use of by chemical engineers from the 1950s through the 1970s seems to have been forgotten or never was learned in some relevant quarters.  Stability theory and mathematics in the context of thermoreactive systems has nothing to do with steady state thermal balances and everything to do with protecting human lives by assaying thermal runaway conditions correctly and accurately. 
 
 We summarize the message of this communication as follows: 
\begin{itemize}
\item It is incorrect, dangerous, and inconsistent with the existing literature on this subject  to assume that thermal runaway is governed only by classical ignition, and use a steady state enthalpy balance to answer   the \textit{wrong} question:  What are the boundaries of classical ignition and extinction points over the parameter space? 
\item The current study has shown that 
\begin{itemize}
\item classical ignition need not be accompanied by thermal runaway,
\item thermal runaway may be non-classical, with ignition to an oscillatory state,
\item non-classical thermal extinction may occur from an oscillatory state, 
\item oscillatory thermal runaway may occur in the complete absence of steady state multiplicity, and 
\item oscillatory thermal instability is endemic to these systems; it is real, it is not some numerical or other artefact, it may occur with violent abruptness, it is potentially dangerous.  It must be taken into account in devising thermal runaway or safe operational criteria. 
\end{itemize}
\item  The \textit{right} question to ask of open  thermoreactive systems is Question 1: Given a steady state, will a small perturbation grow or decay? Answering this question necessarily involves the dynamics; i.e., a stability analysis using the well-founded mathematics of stability theory. 

\end{itemize} 
In view of these results, we strongly recommend that numerical stability analyses be carried out, over the relevant system and operating parameter space, for all exothermic reactions that are run in a CSTR, using procedures that have been  in the published literature  for many decades.

\appendix
\section{}\label{A}
 \begin{table}[H]\caption{\label{table1} Notation and definitions.}
\footnotesize\centerline{\begin{tabular}{p{.1\textwidth}p{.5\textwidth}}
\hline\hline
Symbol or quantity & Definition \\
\hline
$A$ & pre-exponential factor \\
$c$ & concentration of reactant (glycerol or hexamine, mol/l) \\
$C_p$ &molar heat capacity   (J/(mol\,K)\\
$\overline{C}_\text{vol}$ & average volumetric specific heat (J\,l$^{-1}$K$^{-1}$)\\
$\Delta C_p$ & reaction heat capacity  (J/(mol\,K)\\
$E$&activation energy  (J/mol)\\
$F$ & volumetric inflow rate ( l/min) \\
$H$ & enthalpy of formation (J/mol)\\
$\Delta H$ & reaction enthalpy  (J/mol) \\
$k$ & $k(T)$, temperature-dependent rate constant\\
$L$ & combined heat transfer coefficient (W/K)  \\
$n$, $m$ & reaction orders\\
$t$ & time (min)  \\
$T$ & temperature (K) \\ 
$V$ & volume of reaction mixture (l) \\
$x$ & fractional conversion of reactant \\
\hline
\multicolumn{2}{l}{Subscripts, superscripts, overscripts}  \\
\hline
0 & initial value of quantity\\
${}\circ$ & standard state\\
$\dot{}$ & time-changing quantity\\
$\text{a}$ & of the coolant \\
A, B, S & of reactant A, B, of solvent S\\
$\text{f}$ & of the feed stream\\
G & of glycerol\\
N & of nitric acid\\
\hline\hline
\end{tabular}
}
\end{table}

\section{}\label{B} 

The enthalpy balance appears to have been given incorrectly in \cite{Lu:2005} and \cite{Lu:2008}, with the reaction enthalpy $\Delta H^\circ$ being counted twice. The correct enthalpy balance for the adiabatic CSTR housing a first order  exothermic reaction A $\rightarrow$ B in a solvent S was given in \cite{Ball:1999}; it simply expresses the law of energy conservation and is reproduced here for convenient reference:
\begin{multline}\label{a1}
V\frac{\d}{\d t} \left[c_\text{A}H_\text{A}(T) + \left(c_\text{f}-c_\text{A}\right)H_\text{B}(T)+ c_\text{S}H_\text{S}(T)\right] =\\
F\left[c_\text{f}H_\text{A}(T_\text{f})-c_\text{A}H_\text{A}(T)- \left(c_\text{f}-c\right)H_\text{B}(T)\right] \\+ F\left[c_\text{S,f}H_\text{S}(T_\text{f})- c_\text{S}H_\text{S}(T) \right]
 \end{multline} 
Using the thermodynamic relation $C_p=(\partial H/\partial T)_p$  Eq. \ref{e2} may be developed into 
\begin{equation}\label{a2} 
V\frac{\d}{\d t} \left[\overline{C}T +\left(-\Delta H\right)c_\text{A}\right] = -F\left[\left(\overline{C}T+\left(-\Delta H\right)c_\text{A}\right) - \left(\overline{C}T_\text{f}+\left(-\Delta H\right)c_\text{f}\right) \right]. 
 \end{equation} 
 The differential mass balance
 \begin{equation}
 V\frac{\d c_\text{A}}{\d t} = -Vk(T)c_\text{A} + F(c_\text{f} - c_\text{A})
 \end{equation} 
 is substituted into Eq. \ref{a2} to give,  for arbitrary initial conditions, 
  \begin{equation}\label{a4} 
 V\overline{C}\frac{\d T}{\d t} = V(-\Delta H) k(T) c_\text{A} - F\overline{C}(T-T_\text{f}). 
  \end{equation} 
  One may include any number of reaction components and heat loss terms in  the fundamental energy conservation expression Eq \ref{a1}. The correct enthalpy balance, Eq. \ref{a4} emerges naturally by considering the dynamics; i.e., the rate of enthalpy generation, the left hand side of Eq. \ref{a1}.

\section*{Acknowledgement} 
 R. Ball is recipient of Australian Research Council Future Fellowship FT0991007. 
 
% \bibliographystyle{elsarticle-harv.bst}
% \bibliography{/Users/rowena/bib/Thermalrunaway.bib}

\end{document}